\documentclass[conference]{IEEEtran}
\IEEEoverridecommandlockouts
\usepackage{algorithm}
\usepackage{array}
\usepackage[caption=false,font=normalsize,labelfont=sf,textfont=sf]{subfig}
\usepackage{url}
\usepackage{verbatim}
\usepackage{tikz}
\usepackage{circuitikz}
\usetikzlibrary{positioning}

\usepackage{cite}
\usepackage{amsmath,amssymb,amsfonts}
\usepackage{algorithmic}
\usepackage{graphicx}
\usepackage{textcomp}
\usepackage{xcolor}
\def\BibTeX{{\rm B\kern-.05em{\sc i\kern-.025em b}\kern-.08em
    T\kern-.1667em\lower.7ex\hbox{E}\kern-.125emX}}
    
\begin{document}

\title{Enhanced Entropy-Based Metric for Characterization of Delayed Voltage Recovery}
\author{
\IEEEauthorblockN{Mohammad Almomani, Muhammad Sarwar, and Venkataramana Ajjarapu}
\IEEEauthorblockA{\textit{Department of Electrical \& Computer Engineering, Iowa State University, Ames, IA, USA} \\
Emails: \{mmomani, msarwar, vajjarap\}@iastate.edu}
}

\maketitle

\begin{abstract}
Ensuring accurate violation detection in power systems is paramount for operational reliability. This paper introduces an enhanced voltage recovery violation index (EVRVI), a comprehensive index designed to quantify fault-induced delayed voltage recovery (FIDVR). EVRVI enhances traditional entropy-based methods by leveraging Empirical Mode Decomposition (EMD) to extract key features from the voltage signal, which are then used to quantify over-voltage (OV) and under-voltage (UV) events.  Our simulations on the Nordic system, involving over 245k scenarios, demonstrate EVRVI's superior ability to identify and categorize voltage recovery issues compared to the traditional entropy based measure. EVRVI not only significantly reduces false negatives in violation detection but also provides a reliable framework for over-voltage detection, making it an invaluable tool for modern power system studies.

\end{abstract}


\section{Introduction}

\textcolor{black}{FIDVR refers to a phenomenon where, following a fault, the voltage in a power system does not recover to its nominal level within the expected timeframe, often exhibiting sustained low-voltage levels or oscillatory behavior. This delayed recovery is particularly problematic in modern power systems with a high penetration of induction motor loads and renewable energy sources \cite{fundamental}, as it can lead to equipment malfunctions, cascading outages, and compromised grid stability. Identifying and quantifying FIDVR events is essential to mitigate their impact, as these events challenge traditional voltage stability metrics and necessitate more accurate indices for detection and analysis. By addressing these complexities, methods for FIDVR characterization enable more robust power system planning and operation, ensuring enhanced reliability and resilience under fault conditions.}

The entropy-based method using Kullback-Leibler (KL) divergence \cite{KL1} is commonly used to quantify FIDVR by comparing deviations of post-fault voltage signals from a reference. While effective for basic measurements, this method faces challenges in handling scenarios with oscillations or over-voltages, complicating accurate FIDVR quantification. These limitations highlight the need to refine the KL divergence approach to better capture both pure FIDVR events and those influenced by oscillatory or over-voltage conditions. 

EMD emerges as a powerful tool in this context, known for its effectiveness in decomposing non-linear and non-stationary signals into intrinsic mode functions. Applying EMD enables a more detailed analysis of the signal's characteristics, allowing the KL-divergence of each monotonic envelope to be calculated for precise quantification of its recovery. This approach provides an accurate index for FIDVR quantification.

\subsection{Literature Review and Research Gap}

The research on delayed voltage recovery has led to various indices that quantify system performance during transient events. In \cite{P1}, the transient voltage dip acceptability index was introduced to assess the severity of voltage dips and their duration. This approach enables the system operator to minimize critical load shedding and maintain transient stability. In \cite{P2}, the transient voltage severity index was introduced, which globally evaluates system performance by averaging local transient voltage deviation indices at each bus. While this method provides a comprehensive system-level overview, it may overlook regional variations in bus behavior under severe disturbances. Numerous studies have refined these indices to improve the accuracy of delayed voltage recovery assessments. Generally, methods for quantifying FIDVR in the literature fall into two categories \cite{KL1}:
\paragraph{ Slope-based methods:} These metrics, which rely on the slope or derivative of voltage progression, are less suitable when voltage exhibits oscillations or abrupt (discontinuous) changes.
 \paragraph{ Integral error-based methods:
  }While these techniques measure the deviation over time, they fail to differentiate between two scenarios: (i) a waveform with a smaller initial voltage drop but slower recovery, and (ii) a waveform with a larger initial drop that recovers quickly within a short time frame.

To deal with the aforementioned limitations, \cite{P3} introduced a contingency severity index that accounts for both the magnitude and timing of voltage limit violations. However, it only considers the most critical points of failure rather than the entire transient period. Reference \cite{KL1} proposed entropy-based index using KL divergence, which takes a probabilistic approach to characterizing voltage recovery by comparing observed waveforms to reference performance metrics. This index improves upon earlier approaches by providing a more detailed understanding of voltage recovery patterns during disturbances.

However, this method faces challenges in addressing delayed recovery accompanied by oscillations, a limitation that the voltage recovery index (VRI), introduced in \cite{P5}, seeks to overcome. VRI incorporates weighting functions that reward or penalize voltage recovery to enhance the KL divergence method, helping it better quantify oscillatory recovery. VRI was developed based on the observation that the KL divergence can yield a higher index when the probability at 1 pu exceeds the normal distribution value at 1, as illustrated in \cite{P5}. The gap in understanding here is that this issue can be managed by decreasing the standard deviation of the normal distribution. However, while VRI improves the index by penalizing oscillations, it does not specifically address oscillations occurring during FIDVR events.

Based on \cite{P5}, \cite{P6} proposed a new update for entropy-based index to assess the voltage recovery at the system level (VRI$_{sys}$), which enhances the VRI by creating a global measure of voltage recovery across all buses in a power system. This index is based on a weighted average of the VRI values at each bus, considering the electrical distance from the fault location and voltage recovery at different buses. VRI$_{sys}$ provides a complete assessment of short-term voltage assessment, especially for large-scale systems with significant non-conventional generation penetration, such as wind and solar. The global index in \cite{P6} leverages the VRI from \cite{P5}; However, it does not effectively address over-voltage issues or accurately manage oscillations that occur during FIDVR events.

\subsection{Contribution of This Paper}
 This paper introduces the EVRVI and addresses several key limitations of the previous work on entropy-based KL divergence measure. The primary limitations of the KL measure that EVRVI addresses include:

\paragraph{ \textbf{Over-voltage and Under-voltage Detection:}}The KL divergence focuses on detecting deviations from reference signals but does not efficiently capture cases of over-voltage or under-voltage, which can be equally harmful to power system stability. EVRVI is designed to detect both under-voltage and over-voltage violations, providing a more comprehensive assessment of voltage recovery.

\paragraph{\textbf{Managing Oscillations During Recovery:}} While the KL measure is sensitive to deviations, it cannot effectively handle oscillations during recovery. To address this, EVRVI introduces an additional criterion that minimizes the impact of oscillations on the KL divergence, allowing a focus on smooth, monotonic recovery behavior.

\paragraph{\textbf{Enhanced Comparison of Voltage Signals:} }Traditional KL measures are often limited when comparing voltage signals with varying profiles. The EVRVI framework improves accuracy by enabling a more adaptable comparison of similar distribution functions, allowing for a more consistent and precise evaluation of voltage recovery behaviors and overall system stability.

\section{Background}

\subsection{KL Divergence Measure }

KL divergence, also known as relative entropy, is a fundamental concept in information theory introduced by Solomon Kullback and Richard Leibler in 1951. It quantifies the difference between two probability distributions, \(P\) and \(Q\). Mathematically, KL divergence is defined as:
\begin{equation}
\centering
   D_{KL}(P||Q) = \sum_i P(i) \log \frac{P(i)}{Q(i)}
\label{eq:KL} 
\end{equation}
for discrete distributions, and as an integral for continuous distributions. This measure is non-symmetric and always non-negative, with \(D_{KL}(P||Q) = 0\) if and only if \(P\) and \(Q\) are identical. 

In the context of FIDVR, KL divergence can quantify the deviation of post-fault voltage profiles from the ideal behavior, denoted as \(Q=P^{Ideal}\), providing a rigorous measure for detecting and analyzing FIDVR events. 
To calculate the entropy-based index for a voltage signal, the voltage axis is divided into \( N \) segments, and the sample count in each segment is normalized to obtain the mass probability density function (mdf) of the voltage profile. Similarly, the mdf of a normal distribution over the same range is obtained. The KL divergence then measures the statistical distance between the voltage profile’s mdf and the ideal normal distribution \( P^{Ideal} \) using \ref{eq:KL}. A smaller KL value indicates a profile closer to the normal distribution, reflecting good recovery behavior, while a higher KL value indicates poorer recovery.

\begin{figure}
    \centering
    \includegraphics[width=0.9\linewidth]{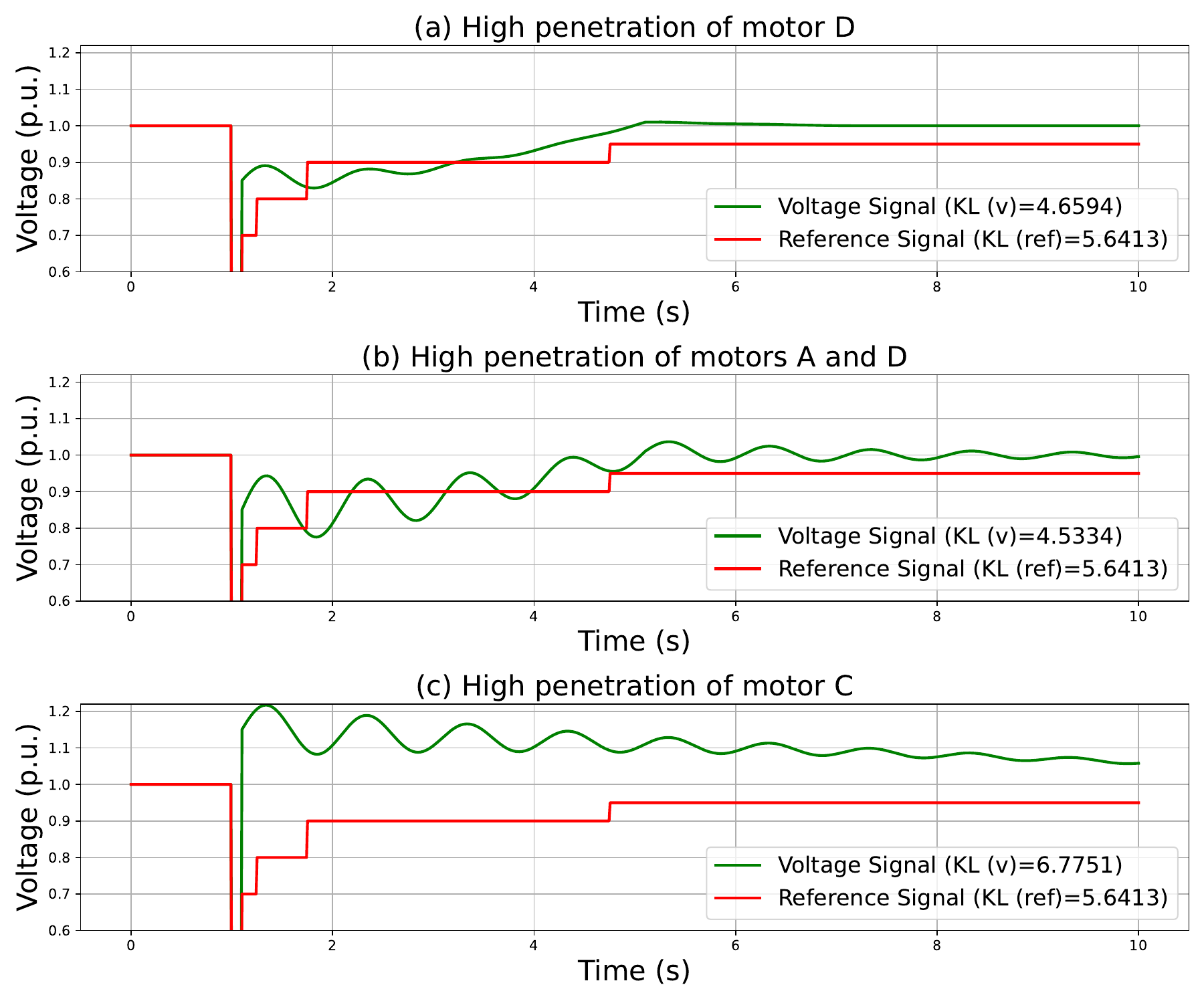}
    \caption{False Detection cases using KL Divergence}
    \label{fig:KLVIOLATION}
\end{figure}

A voltage violation criterion defines the minimum acceptable voltage level. This work adopts the WECC voltage violation criterion \cite{WECC}, used in various studies (\cite{KL1, part1, sarwar}) to evaluate FIDVR. Similarly, Exelon in PJM has defined another stepwise criteria in \cite{PJM}. Figure \ref{fig:KLVIOLATION} shows an example of a stepwise reference voltage criterion with three sample recovery curves. A violation is detected if the KL divergence of the voltage signal \( KL(v) \) exceeds that of the reference signal \( KL(\text{ref}) \), indicating the voltage has crossed the acceptable threshold.

KL divergence can sometimes lead to false detections, identifying oscillations, over-voltage, or smooth recovery as violations (false positives) or missing actual violations (false negatives). To examine this, we use a composite load model with four motor types: Motor A (low inertia, constant torque), Motor B (high inertia, quadratic torque), Motor C (low inertia, quadratic torque), and Motor D (single-phase HVAC). Motor A causes oscillations, Motor B over-voltage, Motor C both over-voltage and oscillations, and Motor D mainly FIDVR events. These dynamics complicate the accuracy of KL divergence, as it encounters difficulty distinguishing these conditions across motor types.

 Figure \ref{fig:KLVIOLATION} illustrates three cases where the traditional entropy-based measure encounters difficulty in accurately detecting violations in voltage recovery profiles. Despite the values of parameters $\lambda$ and $N$ being selected according to the recommended guidelines \cite{KL1}, the results reveal the sensitivity of the KL measure to these parameters, as well as its limitations when dealing with different types of voltage recovery behaviors.

In the \textbf{(a) High penetration of motor D}, we observe a case where two different signal profiles are compared. The KL divergence fails to detect a violation, as the recovered signal exhibits a different shape from the stepwise reference signal, yet the KL value is lower than the threshold.  In the case of high penetration of both motor A and D \textbf{(Figure 1 (b))}, we see the impact of oscillations on the KL measure. The oscillatory nature of the recovered signal results in a KL value that is lower than that of the reference signal, despite the clear violation of the recovery profile. Moreover, the \textbf{(c) High penetration of motor C} illustrates an over-voltage condition where the KL measure detects a violation. This is another drawback of the KL method, as it was originally designed to detect voltage dips, not over-voltage conditions, leading to false positives.

\subsection{Empirical Mode Decomposition}

EMD is a powerful, adaptive signal processing technique introduced by N. E. Huang and colleagues in 1998 \cite{EMD8}. It is specifically designed to analyze non-linear and non-stationary time series data by decomposing the original signal into a set of intrinsic mode functions (IMFs) and a residual. Each IMF represents a simple oscillatory mode, capturing different frequency components of the signal. This process is analogous to an adaptive wavelet transform but without requiring predefined basis functions. The flexibility and data-driven nature of EMD make it particularly suitable for our application. The computational process of EMD involves several iterative steps \cite{wang2010intrinsic}:

\begin{enumerate}
    \item  Identify all the local maxima and minima of the signal. 
    \item Interpolate the local extrema  to form an upper envelope \(U(t)\) and a lower envelope \( L(t)\).
    \item  Calculate the mean of the upper and lower envelopes.
    \item Subtract the mean envelope from the original signal to produce a new signal. This new signal is considered an IMF if it satisfies two conditions: (1) the number of extrema and zero crossings must either be equal or differ by one, and (2) at any point, the mean of the envelope should be zero. If these conditions are not met, the sifting process is repeated on the resulting signal.
    \item Once an IMF is obtained, it is subtracted from the original signal to produce a residue. This residue is then subjected to the same process to extract further IMFs.
    \item The process is repeated iteratively on the residue until the residue becomes a monotonic function.
\end{enumerate}

The final residual component, \( r(t) \), represents the overall trend of the original signal after all oscillatory modes are extracted. In the proposed index, EMD is applied to decompose voltage signals, capturing over-voltage and under-voltage separately. The monotonically decreasing upper envelope \( U(t) \) and monotonically increasing lower envelope \( L(t) \) are constructed from the decomposed components as follows:

{\small
\begin{equation}
U(t) = \max \left( 1, r(t) + \max \left( \sum_i \text{IMF}_i \right) \right)    
\end{equation}}
{\small
\begin{equation}
L(t) = \min \left( 1, r(t) + \min \left( \sum_i \text{IMF}_i \right) \right)    
\end{equation}}

By defining the upper and lower envelopes in this way, the EMD framework effectively separates over-voltage and under-voltage cases. Using \( U(t) \) as a monotonically decreasing function and \( L(t) \) as a monotonically increasing function enables a structured approach to assessing voltage recovery. Deviations from these monotonic envelopes highlight potential delayed recoveries, offering insights into the signal's behavior during recovery phases. Additionally, these monotonic functions are designed to follow a similar distribution, making the KL divergence a more meaningful statistical measure for comparing different voltage recovery trajectories to the reference. This alignment in distribution increases the reliability of KL divergence as a metric for quantifying deviations in recovery performance.

\section{Enhanced Index Using EMD}

The extended entropy-based index combines two
statistical measures to assess delayed voltage recovery events. The index leverages KL divergence to quantify over-voltage and under-voltage recovery from EMD envelope results.

The first component of the index evaluates over-voltage by computing the KL divergence between the monotonically decreasing upper envelope distribution of the voltage  signal and an ideal distribution, represented by a normal distribution centered at 1. This divergence quantifies how closely the actual voltage  path aligns with the ideal profile, with lower KL values indicating a more effective recovery.

Similarly, the second component of the index evaluates under-voltage  by computing the KL divergence between the monotonically increasing lower envelope distribution of the voltage  signal and the same ideal normal distribution centered at 1. Similar to the first component, with lower KL values signifying improved recovery. This component represents the under-voltage delayed recovery index.
The proposed stability index is defined as:
\begin{equation}
    \begin{split}
        D_{\text{KL}}^{u} = D_{\text{KL}}(P_1 \parallel \mathcal{N}(\mu, s^2)) \\
       D_{\text{KL}}^{l} = D_{\text{KL}}(P_2 \parallel \mathcal{N}(\mu, s^2)) 
    \end{split}
\label{eq:proposed}
\end{equation}
For over-voltage and under-voltage respectively. Where:
\begin{itemize}
    \item \(P_1 \sim L(t) \): Represents the distribution of  the lower envelope \(L(t)\).
    \item \(P_2 \sim U(t) \): Represents the distribution of  the upper envelope \(U(t)\).
    \item \(\mathcal{N}(\mu, s^2)\): The normal distribution with mean \(\mu\) and variance \(s^2\), typically set to \(\mu = 1\) and \(s\) as a small positive value, symbolizing the ideal steady-state voltage level during normal operation conditions.
\end{itemize}

For further analysis within the evaluation framework, it is crucial to establish critical \textbf{threshold values: \(D_{\text{violate}}\).} This threshold is defined based on the KL index relative to a standard reference signals. For the under-voltage recovery threshold \(D_{\text{violate, UV}}\), the upper envelope of the reference's low-voltage criteria serves as the benchmark. Conversely, for the over-voltage scenario, the lower envelope of the reference's high-voltage criteria is employed to establish the violation threshold \(D_{\text{violate, OV}}\).

\subsection{Step-by-Step Implementation}
Voltage recovery assessment involves calculating the KL divergence between the envelope of the voltage recovery signal and an ideal reference distribution modeled as a normal distribution centered at 1. Let \( P_{1}(v_i), P_{2} (v_i) \) represent the discrete probability distribution of the voltage recovery upper/lower envelop at discrete points \( v_i \), respectively. The KL divergence for voltage recovery (from \ref{eq:KL}) is calculated as:
{\small
\begin{equation}
D_{\text{KL}} = - \sum_i \mathbf {H}(P [v_i]) -P [v_i]\left( \frac{(v_i-1)^2}{2s^2}+ \frac{1}{2} log (2 \pi s^2)\right)    
\end{equation}}
Where \(P=P_1\) or \( P_2\)  to find \(D_{\text{KL}}^u\) or \(D_{\text{KL}}^l\) respectively, and the reference ideal distribution \( \mathcal{N}(v_i) \) is modeled as a normal distribution with mean \( \mu = 1 \) and small standard deviation \(s^2\). \(\mathbf{H(x)}=-x \ln (x)\) is the entropy

The EVRVI is defined by two indices: \( \text{EVRVI}^+ \), which measures over-voltage violations, and \( \text{EVRVI}^- \), which captures under-voltage violations. These indices are calculated as a ratio between \( D_{KL}^u \) and \( D_{KL}^l \) to \( D_{\text{violate, OV}} \) and \( D_{\text{violate, UV}} \), respectively. These indices allow for a direct comparison between the reference signal and any measured voltage signal. 

The calculation process begins by dividing the voltage axis into \( N \) partitions. If \( \Delta T_i \) represents the time the voltage remains in partition \( i \) and the total time is \( T \), then \( \text{EVRVI}^+ \) and \( \text{EVRVI}^- \) are given by:
{\small
\begin{equation}
    \text{EVRVI}(x) = \frac{\log \left( \frac{Z}{T} \right) + \sum_i \frac{\Delta T_i}{T} \left( \log(\Delta T_i) + \frac{(x_i - 1)^2}{2s^2} \right)}{\log \left( \frac{Z}{T} \right) + \sum_i \frac{\Delta T_i}{T} \left( \log(\Delta T_i) + \frac{(x_{\text{ref},i} - 1)^2}{2s^2} \right)}
\label{EVRVI}
\end{equation}}
Where \( x_i \) represents the values of the upper or lower envelopes in segment \( i \) for \( \text{EVRVI}^+ \) and \( \text{EVRVI}^- \), respectively. Similarly, \( x_{\text{ref},i} \) denotes the values of the upper or lower reference envelopes in segment \( i \) for \( \text{EVRVI}^+ \) and \( \text{EVRVI}^- \), respectively, and \( Z \) is given by \(
 Z = \int_X e^{-\frac{(x - 1)^2}{2s^2}} dx
\)

\section{Simulation Results}

\begin{figure*} 
    \centering
    \includegraphics[width=0.95\textwidth]{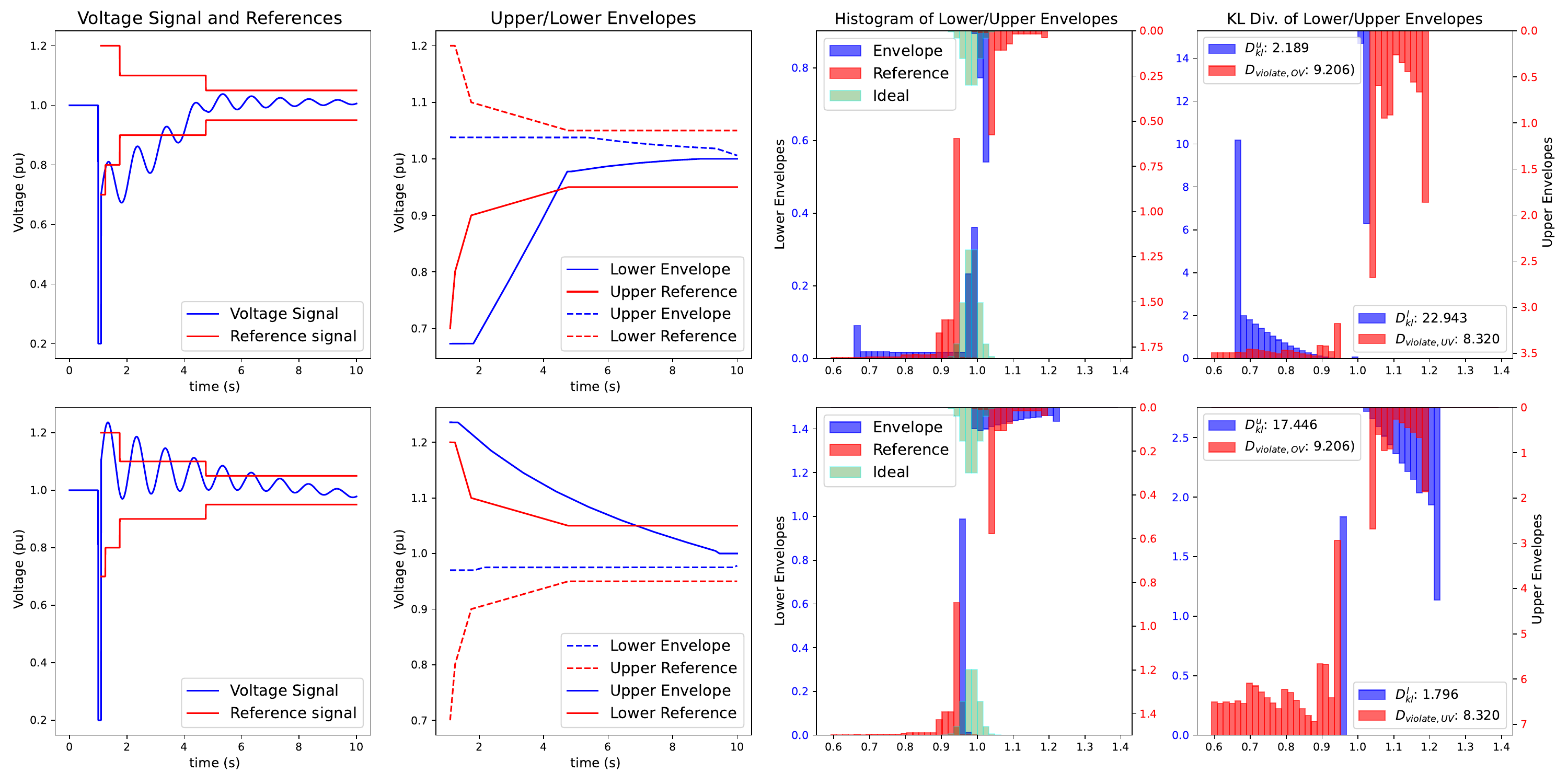}
\caption{\textcolor{black}{ Comprehensive analysis of delayed voltage recovery signals using the proposed extended index. The figure illustrates two cases: under-voltage (top row) and over-voltage (bottom row). The columns depict (1) voltage signals and stepwise reference criteria, (2) lower and upper envelopes calculated using equations (2) and (3), (3) probability distributions of the envelopes for \(N\) segments, and (4) the EVRVI calculated using equations (5) and (6).}}
    \label{fig:stabel_case}
\end{figure*}

The simulation was carried out using the Nordic system at operation point A \cite{testSystems}. The composite load model parameters for AC motors were based on the details provided in \cite{fundamental}. For the KL divergence measure, the simulation was run for 10 seconds to capture the voltage behavior.
\textcolor{black}{A total of \textbf{245,729} scenarios were analyzed (These scenarios are discussed in \cite{fundamental}), with \textbf{216,477} non-violation cases and \textbf{29,252} violation cases, where signals exceeded the reference by at least \textbf{0.005 pu for 250 ms or more}. The KL measure correctly classified \textbf{215,799} non-violation cases with \textbf{678} false positives (mainly over-voltage cases). For violations, it detected \textbf{17,222} true positives, but \textbf{12,030} cases were misclassified as false negatives, resulting in a \textbf{false negative rate of 41.1\%}. This indicates a significant number of undetected voltage violations. The overall accuracy was \textbf{58.9\%} for violation cases and \textbf{99.7\%} for non-violations. All cases were accurately classified by the extended index without errors.}
\textcolor{black}{The implementation of EVRVI is illustrated in Figure \ref{fig:stabel_case}.} 
In the first row, the KL divergence values indicate that the lower envelope of the voltage signal exceeds the threshold, suggesting an under-voltage violation, where \( \text{EVRVI}^-\simeq  2.76 \), successfully detecting the violation. Similarly, the second row subplots outline the calculation steps for the over-voltage signal. Here, the KL divergence values indicate an over-voltage violation with \( \text{EVRVI}^+ \simeq 1.8 \).

\textcolor{black}{ To assess the robustness of the proposed EVRVI, we conducted a sensitivity analysis by varying the key parameters that affect traditional KL divergence, including the number of voltage partitions (\(N\)) and the variance (\(\lambda\)) of the normal distribution. Our results indicate that the traditional KL divergence is highly sensitive to these parameters, with the final KL values changing significantly based on the chosen \(N\) and \(\lambda\). For instance, when comparing a signal with the reference, the KL divergence may inconsistently detect a violation for certain values of \(\lambda\) while indicating no violation for others. Similarly, varying \(N\) affects the calculated values, leading to inconsistencies in detecting violations. In contrast, the proposed EVRVI (Equation 6) demonstrated robustness to such parameter variations. The EVRVI does not depend on \(N\), and our sensitivity study across different variances (\(s^2\)) showed that the index remains stable. This is attributed to the fact that \(s^2\) appears in both the numerator and denominator of the EVRVI calculation, effectively canceling out any dependency. Additionally, the use of monotonically increasing or decreasing envelopes ensures that the same signal pattern is consistently compared, further enhancing the reliability of the proposed index under varying conditions. }

\textcolor{black}{The proposed index is applicable to \textit{system performance monitoring} and \textit{FIDVR assessment}, similar to the methods discussed in [2, 3, 5, 7, and 8]. However, it demonstrates higher accuracy in detecting voltage violations. Compared to the traditional KL measure, the proposed index requires slightly more computational time due to the extraction of upper and lower envelopes, increasing computation time by approximately 1.2 times.}

\section{conclusion} 
\small This paper introduced the Enhanced Voltage recovery Violation Index (EVRVI) as a comprehensive index for quantified delayed voltage recovery. Our research effectively addresses the limitations of the traditional KL divergence measure. Through rigorous simulations conducted on the Nordic system, we demonstrated EVRVI's superior accuracy in assessing voltage stability, notably excelling in identifying undervoltage violations and oscillatory behaviors.  Although the traditional KL divergence measure effectively identified non-violation cases with a low false positive rate, it suffered from a substantial false negative rate of 41.1\%, indicating a significant number of undetected voltage violations.
EVRVI represents a substantial advancement over existing methods for voltage stability assessment and violation detection. Future efforts will be directed towards optimizing ESVI's computational efficiency to facilitate its real-time implementation in large-scale power systems.

\section*{Acknowledgment}
{\small This work was supported by the DOE through CyDERMS project (DOE CESER DE-FOA-0002503), the Power System Engineering Research Center (PSERC) and  
 the National Science Foundation (NSF).}

\bibliographystyle{IEEEtran}  
\bibliography{ref}

\end{document}